\documentclass{article}
\usepackage[T1]{fontenc} 
\usepackage[utf8]{inputenc} 
\usepackage{ismir,amsmath,cite,url}
\usepackage{graphicx}
\usepackage{color}
\usepackage{multirow}
\usepackage{booktabs}

\title{Assessing Alignment in Automatic Piano Transcription Datasets: A Benchmark for Scoring Methods}

\multauthor
{Lucas de Souza Brandão$^1$ \hspace{1cm} Lucas Gris$^2$ \hspace{1cm} Joaquim Cavalcante$^3$} { \bfseries{Chukwuemeka Nkama$^4$ \hspace{1cm} Andrea Poltronieri$^4$ \hspace{1cm}}\\
$^1$ Universidade de São Paulo, Brazil
$^2$ Universidade Federal de Goiás, Brazil\\
$^3$ Universidade Federal da Paraíba, Brazil
$^4$ Universitat Pompeu Fabra, Spain\\
{\tt\small lucasouza@usp.br, lucas.gris@discente.ufg.br, joaquim.breno@academico.ufpb.br}
}

\def\authorname{L. Brandão, L. Gris, J. Cavalcante, C. Nkama, and A. Poltronieri}

\usepackage[bookmarks=false,pdfauthor={\authorname},pdfsubject={\papersubject},hidelinks]{hyperref}

\begin{document}

\maketitle

\begin{abstract}

Automatic Music Transcription (AMT) relies on paired audio--MIDI datasets, implicitly assuming near-perfect temporal alignment. However, widely used piano datasets have been shown to contain alignment errors that may bias model evaluation and training. We propose a framework for evaluating audio--MIDI alignment scoring methods through controlled temporal, pitch, and structural MIDI corruptions. We also introduce STI, a scoring method built on AMT models, and benchmark it against existing alignment scoring approaches. Our results show that STI, together with methods based on CENS and spectrograms, achieves the strongest overall performance, with STI exhibiting the highest sensitivity to fine-grained temporal misalignments, the main alignment error in AMT datasets. We then apply the best-performing methods to widely used piano AMT datasets and identify trends in their relative alignment quality. These results support more informed dataset curation, model training, and AMT evaluation.

\end{abstract}

\section{Introduction}\label{sec:introduction}

Automatic Music Transcription (AMT) generally relies on datasets composed of paired audio and MIDI recordings \cite{survey-simonDixon, maestro}. These pairs are intended to represent a perfect correspondence, with all note pitches, onsets, and offsets accurately aligned. Such alignment is fundamental to both training and evaluating AMT systems \cite{onsets-and-frames}, whose objective is to accurately predict these note-level events. However, this assumption has been shown to be inaccurate.

Several studies have reported temporal alignment issues in widely used AMT datasets. By evaluating transcription performance under different onset tolerance windows, \cite{cogliati} and \cite{mt3} found evidence of substantial misalignments in MAPS \cite{maps} and MusicNet \cite{musicnet}, respectively. These findings for MAPS are further supported by \cite{transkun}, which reported improved transcription performance with alternative MAPS annotations and higher performance on SMD \cite{smd}, suggesting that SMD is better aligned than the original MAPS annotations. A more targeted analysis of MAPS identified poorly annotated notes \cite{gong2019analysis}, but focused on a single subset and relied on manual validation. Despite this evidence, there has been little systematic investigation of the extent and severity of alignment errors across piano AMT datasets, with existing analysis often relying on AMT model performance, which can vary for reasons unrelated to alignment.

In this work, we investigate audio--MIDI alignment in automatic piano transcription datasets. We first propose a framework to assess and quantify alignment quality by benchmarking several scoring mechanisms and identifying those best suited to the task. We also introduce a scoring method based on transcription models that is specifically designed to capture fine-grained temporal misalignments at note onsets. The evaluation is based on the response of each method to controlled MIDI corruptions, using original audio--MIDI pairs from standard AMT datasets as references. Finally, we apply the best-performing methods identified by the framework to assess the alignment quality of widely used piano AMT datasets, investigating both the methods' ability to distinguish alignment differences and the alignment quality of the datasets themselves. \footnote{Implementation available on our \href{https://github.com/sb-lucas/SyncAM}{\underline{GitHub}.}}
\section{Method}

We perform controlled corruptions over audio--MIDI pairs, define a set of alignment scoring methods and propose evaluation metrics to assess score behavior.

\subsection{MIDI Corruption}

We construct a pipeline for three types of corruption applied to audio--MIDI pairs: pitch, note onset, and structural. We do not evaluate note velocity alignment because it lacks a well-defined MIDI-to-audio mapping, making controlled corruption of this dimension methodologically unclear within our framework. Each corruption is parameterized by a severity $S \in [0,1]$, discretized into $N=11$ equally spaced levels $S \in \{0, 0.1, \dots, 1\}$, where $S=0$ denotes no corruption and $S=1$ the maximum corruption allowed by the method. The mapping from $S$ to corruption magnitude varies by type, as described below:

\textbf{Pitch Swap:} This corruption modifies note pitches while preserving their original onset times and durations. Each new pitch is sampled uniformly from the standard piano range ([21, 108]), excluding the note's original pitch. The number of swapped notes, $N_{\text{swap}}$, is determined by the severity level $S$ and the total number of notes $N_{\text{orig}}$ in the original MIDI file:

\begin{equation}
\label{severity-pitch}
N_{\text{swap}} = \lfloor S \cdot N_{\text{orig}}\rfloor
\end{equation}

\textbf{Structural:} This corruption randomly removes existing notes and inserts an equal number of new notes into the MIDI file. Added notes are assigned random pitches between 21 and 108 and random velocities between 60 and 100. Each added note is also assigned a random onset time within the range of the original MIDI file and a duration between 0.1~s and 1.5~s. The total number of modified notes, $N_{\text{modified}}$, is the sum of deleted and inserted notes:

\begin{equation}
\label{severity-structural}
N_{\text{modified}} = \lfloor2 \cdot S \cdot N_{\text{orig}} \rfloor
\end{equation}

\textbf{Onset Shift:} This corruption shifts the onset of each note by a value $\Delta t$, randomly advancing or delaying each note by that amount. According to MIREX standards \cite{mirex}, a note is considered correctly aligned if its onset is within 50\,ms of the reference. We therefore set the maximum shift, $\Delta t_{\max}$, to 500\,ms to cover both fine-grained timing errors and coarse temporal misalignments. To better analyze fine-grained misalignments below 50\,ms, we apply a quadratic mapping that spans the full range while ensuring a minimum shift of 5\,ms for any non-zero severity level and allocating four severity levels below 50\,ms. The onset shift is therefore defined as:

\begin{equation}
\label{onset-shift-deltat}
\Delta t = \Delta t_{\max} \cdot S^2
\end{equation}

\subsection{Scores}

We generate scores ranging from 0 to 1, rescaling whenever needed, using four families of methods: embedding-based, feature-alignment-based, onset-based, and transcription-based scores. 

Embedding-based methods compute cosine similarity between audio and MIDI embeddings in a shared latent space. We evaluate \textbf{CLaMP 3} \cite{clamp3}, which directly embeds audio and MIDI into a shared representation, and \textbf{CLAP} \cite{clap}, which first synthesizes the MIDI into audio before comparing audio embeddings. For inputs longer than the model's maximum duration, we split the audio into non-overlapping chunks, compute embeddings independently, and average them.

Feature-alignment-based methods compare audio and MIDI using three frame-level feature representations: spectrograms, CENS \cite{cens}, and DLNCO \cite{dlnco}. We synthesize the MIDI into audio and extract spectrograms and CENS from both the original and synthesized audio. DLNCO is extracted directly from the MIDI and original audio. We evaluate two alignment strategies. The first, based on DTW, uses MrMsDTW~\cite{mrmsdtw} to compute an optimal alignment path between feature sequences and scores the mean frame-wise cosine similarity along that path. We evaluate this strategy on CENS and DLNCO features (\textbf{DTW-C} and \textbf{DTW-D}, respectively). The second uses frame-wise cosine similarity, obtained by L2-normalizing each sequence independently, zero-padding the shorter sequence at the end to match the length of the longer one, and averaging the resulting per-frame cosine similarities over the full padded length. We evaluate this strategy on spectrograms, DLNCO and CENS features (\textbf{CS-S}, \textbf{CS-D} and \textbf{CS-C}, respectively).

Leveraging note onset information, \textbf{Onset-F1} compares MIDI note onsets with those detected from audio using Madmom's CNN-based onset detector \cite{madmom}. MIDI onsets within 30\,ms of each other are merged to simulate chords, matching the behavior of audio onset detectors. We report the mir\_eval~\cite{mir_eval} onset F1-score using the standard MIREX tolerance of 50\,ms.

Transcription-based methods first transcribe audio to MIDI and then score the resulting transcription against the reference MIDI under two matching strategies. The first, following the standard MIREX evaluation protocol, computes a note-level F-measure in which a note is considered correct if its pitch falls within a quarter tone, its onset within 50\,ms, and its offset within 20\% of its duration. We evaluate this strategy using Basic Pitch~\cite{basic-pitch}, Transkun~\cite{transkun}, and
ByteDance~\cite{byte-dance}, resulting in \textbf{AMT-BP}, \textbf{AMT-TK} and \textbf{AMT-BD}, respectively. 

Building on the same models, we propose a novel audio--MIDI alignment scoring method---the \textbf{S}oft, \textbf{T}ranscription-based, pitch-constrained \textbf{I}nter-Onset Interval (\textbf{STI}) score. Each note is represented by its pitch, onset time, and pitch-constrained Inter-Onset Interval (IOI), defined as the time elapsed since the previous note of the same pitch. This representation provides additional structural information for robust MIDI comparison. 

We construct a bipartite cost matrix between the predicted and reference MIDIs. Matching-pitch note pairs receive a cost $c$ given by the equally weighted sum of their absolute onset and IOI differences, while mismatched-pitch pairs receive a large penalty of $10^6$. The globally optimal one-to-one assignment is computed using the Hungarian algorithm. Each matched pair receives a soft weight

\begin{equation}\label{eq:ioi}
w = \exp(-c/\tau),
\end{equation}

where $\tau=0.05$ is the tolerance scale, matching the temporal tolerance used in MIREX. The sum of these weights yields a soft note-match count, from which we compute continuous precision, recall, and F-measure. We evaluate STI with the same three transcription models, resulting in \textbf{STI-BP}, \textbf{STI-TK}, and \textbf{STI-BD}.

\begin{table*}[ht]
\centering
\small
\setlength{\tabcolsep}{6pt}
\renewcommand{\arraystretch}{1.15}
\begin{tabular}{l cccc cccc cccc}
  \toprule
  & \multicolumn{4}{c}{\textbf{Structural}} & \multicolumn{4}{c}{\textbf{Onset Shift}} & \multicolumn{4}{c}{\textbf{Pitch Swap}} \\
  \cmidrule(lr){2-5} \cmidrule(lr){6-9} \cmidrule(lr){10-13}
  \textbf{Metric} & Mono & Stab & Sens & Cons & Mono & Stab & Sens & Cons & Mono & Stab & Sens & Cons \\
  \midrule
  CLaMP 3          & 0.89 & \textbf{0.97} & 0.09 & 0.83 & 0.55 & 0.96 & 0.02 & 0.46 & 0.93 & \textbf{0.96} & 0.07 & 0.87 \\
  CLAP            & 0.77 & 0.96 & 0.03 & 0.69 & 0.25 & 0.96 & 0.00 & 0.26 & 0.22 & \textbf{0.96} & 0.00 & 0.41 \\
  CS-C         & 0.99 & 0.93 & 0.49 & 0.87 & 0.97 & 0.95 & 0.23 & \textbf{0.95} & \textbf{1.00} & 0.94 & 0.49 & 0.99 \\
  CS-D        & 0.98 & 0.68 & 0.78 & 0.98 & 0.72 & 0.65 & 0.66 & 0.42 & 0.99 & 0.69 & 0.75 & 0.98 \\
  CS-S         & 0.99 & 0.93 & 0.34 & 0.87 & 0.97 & 0.94 & 0.21 & 0.89 & \textbf{1.00} & 0.93 & 0.37 & \textbf{1.00} \\
  DTW-C    & 0.96 & 0.94 & 0.45 & 0.83 & \textbf{0.99} & \textbf{0.98} & 0.16 & 0.93 & \textbf{1.00} & 0.95 & 0.45 & 0.98 \\
  DTW-D   & 0.99 & 0.72 & 0.58 & 0.98 & 0.91 & 0.73 & 0.23 & 0.71 & \textbf{1.00} & 0.71 & 0.65 & 0.99 \\
  Onset-F1        & 0.95 & 0.85 & 0.47 & 0.97 & 0.57 & 0.76 & 0.43 & 0.33 & 0.00 & 0.84 & 0.00 & 0.00 \\
  AMT-BP          & 0.98 & 0.56 & 0.93 & 0.98 & 0.61 & 0.52 & 0.74 & 0.45 & 0.99 & 0.54 & 0.88 & 0.98 \\
  AMT-BD          & 0.98 & 0.59 & 0.95 & 0.98 & 0.64 & 0.52 & 0.76 & 0.47 & 0.99 & 0.58 & 0.90 & 0.98 \\
  AMT-TK          & \textbf{1.00} & 0.74 & 0.99 & \textbf{1.00} & 0.64 & 0.62 & 0.80 & 0.47 & \textbf{1.00} & 0.75 & 0.91 & \textbf{1.00} \\
  STI-BP      & \textbf{1.00} & 0.70 & \textbf{1.00} & 0.91 & 0.95 & 0.71 & 0.93 & 0.65 & \textbf{1.00} & 0.71 & \textbf{1.00} & 0.91 \\
  STI-BD      & \textbf{1.00} & 0.76 & \textbf{1.00} & 0.89 & 0.95 & 0.76 & \textbf{0.95} & 0.63 & \textbf{1.00} & 0.76 & \textbf{1.00} & 0.89 \\
  STI-TK      & \textbf{1.00} & 0.76 & \textbf{1.00} & 0.88 & 0.95 & 0.77 & \textbf{0.95} & 0.63 & \textbf{1.00} & 0.77 & \textbf{1.00} & 0.89 \\
  \bottomrule
\end{tabular}
\caption{Performance of each method per corruption type averaged across datasets (Mono: Monotonicity, Stab: Stability, Sens: Sensitivity, Cons: Consistency). Bold marks the best value per column.}
\label{tab:corruption_results}
\end{table*}

\subsection{Evaluation Metrics}
For a fixed corruption type, let $\mathcal{S}=(S_1,\dots,S_N)$ denote the ordered severity levels. A sample $j$ consists of an audio recording and its corresponding corrupted MIDIs for that corruption type. Let $\varphi_{S_i}^{(j)}$ denote the score obtained for sample $j \in \{1,\dots,L\}$ at severity level $S_i$. All metrics are computed independently for each corruption type. When needed, we use $\epsilon$ as a small constant to avoid division by zero. For all, higher values indicate better performance.

\textbf{Monotonicity:} Measures whether scores consistently decrease as severity $S$ increases, regardless of magnitude. Let ${\varphi}^{(j)} = \left(\varphi_{S_1}^{(j)}, \dots, \varphi_{S_N}^{(j)}\right)$ denote the ordered scores for sample $j$. We compute the Spearman rank correlation between $\mathcal{S}$ and ${\varphi}^{(j)}$ for each sample and average across samples. Since the relationship is inverse, we negate it:

\begin{equation}
\text{Monotonicity} = -\frac{1}{L}\sum_{j=1}^{L}\rho_{\text{Spearman}}(\mathcal{S}, \boldsymbol{\varphi}^{(j)})
\end{equation}

\textbf{Stability:} Measures whether scores are stable across samples at a fixed severity level. We compute the normalized dispersion across samples for each severity and average over all severity levels:

\begin{equation}
\text{Stability} = \frac{1}{N}\sum_{i=1}^{N} \frac{1}{1 + \sigma(\varphi_{S_i})/(\mu(\varphi_{S_i})+\epsilon)},
\end{equation}

where $\mu(\varphi_{S_i})$ and $\sigma(\varphi_{S_i})$ denote the mean and standard deviation of $\{\varphi_{S_i}^{(j)}\}_{j=1}^{L}$, respectively.

\textbf{Sensitivity:} Measures the overall responsiveness to corruption, independent of the response shape. It quantifies the relative score drop between $S=0$ and $S=1$ for each sample and averages across samples:

\begin{equation}
\text{Sensitivity} = \frac{1}{L}\sum_{j=1}^{L} \frac{\varphi_{S=0}^{(j)} - \varphi_{S=1}^{(j)}}{\varphi_{S=0}^{(j)} + \epsilon}
\end{equation}

\textbf{Consistency:} Measures whether scores degrade uniformly with corruption magnitude rather than through abrupt transitions. For pitch and structural corruptions, magnitude increases linearly with $S$, so we fit $\varphi = aS+b$ per sample. For onset shift, magnitude follows $\Delta t \propto S^2$ (Eq.~\ref{onset-shift-deltat}), so we fit $\varphi = aS^2+b$. We report the average coefficient of determination $R^2$ per sample:

\begin{equation}
\text{Consistency} = \frac{1}{L}\sum_{j=1}^{L} R^2_{(j)}
\end{equation}

\begin{table*}[ht]
\centering
\small
\setlength{\tabcolsep}{6pt}
\renewcommand{\arraystretch}{1.15}
\begin{tabular}{l cccc cccc cccc}
  \toprule
  & \multicolumn{4}{c}{\textbf{MAESTRO}} & \multicolumn{4}{c}{\textbf{SMD}} & \multicolumn{4}{c}{\textbf{MAPS-S}} \\
  \cmidrule(lr){2-5} \cmidrule(lr){6-9} \cmidrule(lr){10-13}
  \textbf{Metric} & Mono & Stab & Sens & Cons & Mono & Stab & Sens & Cons & Mono & Stab & Sens & Cons \\
  \midrule
  CS-C         & 0.94 & \textbf{0.99} & 0.02 & 0.94 & 0.64 & 0.93 & 0.01 & 0.83 & 0.94 & \textbf{0.99} & 0.02 & 0.91 \\
  CS-S         & 0.95 & 0.97 & 0.04 & 0.97 & 0.75 & 0.92 & 0.02 & 0.88 & 0.97 & 0.96 & 0.04 & 0.94 \\
  DTW-C    & 0.94 & \textbf{0.99} & 0.02 & 0.94 & 0.93 & \textbf{0.99} & 0.02 & 0.93 & 0.95 & \textbf{0.99} & 0.02 & 0.91 \\
  DTW-D   & 0.28 & 0.76 & 0.06 & 0.59 & 0.55 & 0.71 & 0.08 & 0.64 & -0.18 & 0.67 & 0.00 & 0.25 \\
  STI-BP      & \textbf{1.00} & 0.78 & 0.47 & \textbf{1.00} & \textbf{1.00} & 0.63 & 0.34 & \textbf{0.98} & \textbf{1.00} & 0.85 & 0.48 & \textbf{0.99} \\
  STI-BD      & \textbf{1.00} & 0.92 & \textbf{0.52} & 0.99 & 0.98 & 0.66 & \textbf{0.38} & \textbf{0.98} & \textbf{1.00} & 0.96 & \textbf{0.52} & \textbf{0.99} \\
  STI-TK      & \textbf{1.00} & 0.95 & \textbf{0.52} & 0.99 & 0.98 & 0.67 & 0.37 & 0.97 & \textbf{1.00} & 0.98 & \textbf{0.52} & \textbf{0.99} \\
  \bottomrule
\end{tabular}
\caption{Evaluation of well-performing methods on fine-grained onset corruption per dataset (Mono: Monotonicity, Stab: Stability, Sens: Sensitivity, Cons: Consistency). Bold marks the best value per column.}
\label{tab:onset_fine_grained}
\end{table*}

\section{Experiments}

We source audio--MIDI pairs from the validation split of MAESTRO \cite{maestro}, SMD, and what we refer to as MAPS-S, which consists only of musical pieces from the synthetic AkPnCGdD and SptkBGAm subsets of MAPS. This combination provides diversity in timbre and style, enabling assessment of generalization. While MAESTRO and SMD are not perfectly aligned, their errors are typically limited to small temporal misalignments. The perfectly aligned MAPS-S subset, therefore, enables fine-grained evaluation of sensitivity to temporal misalignment.

All metrics use $\epsilon=10^{-8}$ for numerical stability. Embedding-based methods use CLAP \texttt{laion/clap-htsat-unfused} and CLaMP 3 \texttt{sander-wood/clamp3} (SAAS variant). All MIDI-to-audio synthesis use FluidSynth with the FluidR3\_GM soundfont, program 0 (piano). Feature extraction uses a 22,050\,Hz sampling rate and hop length of 64 to extract CENS via Librosa~\cite{librosa}, DLNCO via SyncToolbox \cite{synctoolbox} and Librosa, and magnitude spectrogram features via Librosa with a 512-point Fast Fourier Transform.

\section{Results}

Table~\ref{tab:corruption_results} reports results for all corruption types, averaged over the three datasets. Embedding-based methods consistently show the lowest monotonicity values across all corruption types. In contrast, the proposed STI method substantially outperforms the standard AMT F-measure on onset corruption, demonstrating the benefit of incorporating Inter-Onset Interval information to provide a richer structural representation for MIDI comparison. 

Across corruption types, structural corruption is easily detected by nearly all methods, whereas pitch swaps are either captured accurately or largely missed, exposing clear limitations in some approaches. Onset shifts are considerably more challenging to detect, resulting in a much wider spread of monotonicity values.
 
Since onset shift is the most discriminative corruption type and AMT datasets primarily suffer from fine-grained temporal misalignments, Table~\ref{tab:onset_fine_grained} focuses on the first four onset severity levels, corresponding to shifts of 0\,ms, 5\,ms, 20\,ms, and 45\,ms. Results are reported separately for MAESTRO, SMD, and MAPS-S to evaluate the influence of timbre and recording conditions. Only methods with onset monotonicity above 0.9 in Table~\ref{tab:corruption_results} are included. Comparing the two tables, DTW-D exhibits a drop in monotonicity, suggesting that it captures coarse temporal misalignments better than fine-grained ones. More generally, Table~\ref{tab:onset_fine_grained} shows lower monotonicity on SMD than on MAESTRO and MAPS-S for most methods, suggesting the presence of temporal alignment noise in SMD. 

Methods based on spectrogram and CENS features perform best in stability. Their slowly varying spectral representations make them less sensitive to small onset shifts, resulting in stable but less informative scores. In contrast, the proposed STI methods trade some stability for greater sensitivity and resolution in detecting fine-grained misalignments, achieving the best overall performance by ranking among the top methods across all other metrics.

\begin{table}[t]
\centering
\small
\setlength{\tabcolsep}{2.5pt}
\renewcommand{\arraystretch}{1.15}
\begin{tabular}{l ccccc}
  \toprule
  \textbf{Dataset} & \textbf{CS-S} & \textbf{DTW-C} & \textbf{STI-BP} & \textbf{STI-BD} & \textbf{STI-TK} \\
  \midrule
  SMD      & $71 \pm 6$ & $\underline{92} \pm 1$ & $31 \pm 19$ & $58 \pm 31$ & $61 \pm 32$ \\
  MAESTRO  & $74 \pm 3$ & $91 \pm 1$ & $34 \pm 10$ & $\underline{82} \pm 8$ & $\underline{87} \pm 5$ \\
  MAPS-S   & $\mathbf{78} \pm 3$ & $\mathbf{93} \pm 1$ & $\mathbf{47} \pm 8$ & $\mathbf{90} \pm 4$ & $\mathbf{93} \pm 2$ \\
  ENSTDkAm & $72 \pm 2$ & $89 \pm 1$ & $33 \pm 6$ & $51 \pm 8$ & $67 \pm 8$ \\
  ENSTDkCl & $\underline{75} \pm 2$ & $89 \pm 2$ & $\underline{39} \pm 9$ & $64 \pm 9$ & $73 \pm 8$ \\
  MusicNet & $71 \pm 2$ & $87 \pm 4$ & $25 \pm 8$ & $42 \pm 12$ & $45 \pm 11$ \\
  \bottomrule
\end{tabular}
\caption{Dataset alignment quality according to the methods with highest monotonicity performance ($\text{mean} \pm \text{std}$, rescaled to $[0, 100]$). Bold marks the best performance and underline marks the second-best performance per column.}
\label{tab:dataset_alignment}
\end{table}

We select the five methods with highest monotonicity performance on MAPS-S for fine-grained onset alignment to assess real-world dataset alignment. Table~\ref{tab:dataset_alignment} reports their scores across six piano AMT datasets: the three previously used datasets, the MUS subsets of ENSTDkAm and ENSTDkCl from MAPS, and MusicNet's solo piano recordings. All methods rank the synthetic and perfectly aligned MAPS-S highest, confirming prior expectations. However, they disagree on the alignment quality of recorded datasets, leaving only general trends. Most methods rank ENSTDkCl above ENSTDkAm, while SMD shows the largest score variability, suggesting inconsistent alignment across samples. MAESTRO remains among the three best-aligned datasets but is not consistently ranked first. Lastly, MusicNet is consistently scored as the worst-aligned.

The similar near-ceiling scores produced by CS-S and DTW-C limit discrimination. In contrast, the proposed STI methods provide a wider score range, enabling more effective comparison of fine-grained alignment differences.

\section{Conclusion}\label{sec:conclusion}

In this work, we propose a framework for evaluating audio--MIDI alignment scoring methods through controlled pitch, onset, and structural corruptions applied to piano AMT datasets. We introduce four metrics for different aspects of score behavior and propose STI, a scoring method that leverages modern AMT models. 

Our benchmark identifies feature-based methods and STI as the strongest overall approaches. In particular, STI provides high resolution for fine-grained temporal misalignments, making it well suited to the small onset errors that characterize alignment issues in AMT datasets. We then apply the best-performing methods to rank the alignment quality of widely used piano AMT datasets and find that they produce general consistent rankings, although they disagree on the precise ordering of closely aligned datasets. Our benchmark helps identify and measure the conditions under which methods differ, providing a systematic basis for comparing and selecting alignment scoring methods. Building on these findings, future work includes improving STI as AMT quality advances, extending it to other instruments, and refining the benchmark for broader and more reliable alignment assessment.

\section{Acknowledgments}

This study was financed, in part, by the São Paulo Research Foundation (FAPESP), Brasil. Process Number 2025/02767-4. We also acknowledge the support provided by Rede de Avanço em Inteligência Artificial (RAIA).


\bibliography{references}

\end{document}